\documentclass{pasj00}
\draft

\begin{document}
\SetRunningHead{Author(s) in page-head}{Running Head}
\Received{2000/12/31}
\Accepted{2001/01/01}

\title{Formation of Galactic Center Magnetic Loops}

\author{Mami \textsc{Machida}}%
\affil{Division of Theoretical Astronomy, 
National Astronomical Observatory of Japan, \\
Osawa, Mitaka, Tokyo 181-8588, Japan}
\email{mami@th.nao.ac.jp}

\author{Ryoji \textsc{Matsumoto}}
\affil{Faculty of Science, Chiba University, 
Yayoi-cho, Inage-ku, Chiba, 263-8522, Japan}
\author{Satoshi \textsc{Nozawa}}
\affil{Department of Science, Ibaraki University, 
2-1-1 Bunkyo, Mito, Ibaraki 310-8512, Japan}
\author{Kunio \textsc{Takahashi}}
\affil{CfCA, National Astronomical Observatory of Japan, 
Osawa, Mitaka, Tokyo 181-8588, Japan}
\author{Yasuo \textsc{Fukui}, Natsuko \textsc{Kudo}, 
Kazufumi \textsc{Torii}, Hiroaki \textsc{Yamamoto}, 
Motosuji \textsc{Fujishita}}
\affil{Department of Astrophysics, Nagoya University, 
Chikusa-ku, Nagoya 464-8602, Japan}
\and
\author{ Kohji {\sc Tomisaka}}
\affil{Division of Theoretical Astronomy, 
National Astronomical Observatory of Japan, \\
Osawa, Mitaka, Tokyo 181-8588, Japan}


%

\KeyWords{Galaxy: Magnetic field, MHD, Magnetic loops, ISM}

\maketitle

\begin{abstract}
A survey for the molecular clouds in the Galaxy with NANTEN mm telescope 
has discovered molecular loops in the Galactic center region. 
The loops show monotonic gradients of the line of sight velocity 
along the loops and the large velocity dispersions 
towards their foot points. 
It is suggested that these loops are explained in terms of 
the buoyant rise of magnetic loops 
due to the Parker instability. 
We have carried out global three-dimensional magneto-hydrodynamic 
simulations of the 
gas disk in the Galactic center. 
The gravitational potential is approximated by the axisymmetric 
potential proposed by \citet{miy1975}. 
At the initial state, we assume a warm ($ \sim 10^4{\rm K}$) gas 
torus threaded by azimuthal magnetic fields. 
Self-gravity and radiative cooling of the gas are ignored. 
We found that buoyantly rising magnetic loops are formed 
above the differentially rotating, magnetically turbulent disk. 
By analyzing the results of global MHD simulations, 
we have identified individual loops, about 180 in the 
upper half of the disk, and studied 
their statistical properties such as their length, 
width, height, and velocity distributions along the loops. 
Typical length and height of a loop are 1kpc and 200pc, respectively. 
The line of sight velocity changes linearly along a loop 
and shows large dispersions around the foot-points. 
Numerical results indicate that loops emerge preferentially 
from the region where magnetic pressure is large.
We argue that these properties are consistent with those 
of the molecular loops discovered by NANTEN. 
\end{abstract}

\section{Introduction}
Spiral galaxies generally possess large-scale magnetic fields 
(e.g., \cite{sof1986, bec1996}). 
These magnetic fields are usually considered to be amplified and 
maintained by large-scale dynamo action driven by 
collective inductive effects of turbulence and differential rotation
\citep{par1971}.  
In our Galaxy, the strength of the magnetic field is about 
a few $\mu$ G. 
Inside 200pc from the Galactic center, however, 
radio observations indicate that magnetic fields are as strong 
as a few mG (e.g., \cite{mor1996}). 
Recent observations of other spiral galaxies indicate that 
the typical averaged equipartition strength of the 
total magnetic fields is about $10 \mu$G \citep{bec2008}.

Based on a survey for the molecular gas in the Galaxy by NANTEN
(e.g., \cite{miz2004}). 
\citet{fuk2006} found loop-like 
structures of dense molecular gas showing large line-of-sight 
velocity gradients ($\sim 30 {\rm km/s}$) along the loops and large velocity 
dispersions at their foot-points. 
The total mass of the molecular gas estimated from the observation 
is about $1.7 \times 10^5 M_{\odot}$ when 
LTE at 50K was assumed at a distance of 8.5 kpc. 
The kinetic energy of the loops estimated from their line-of-sight 
velocity gradient exceeds $10^{51}$ ergs.  
This energy exceeds the energy injected by a supernova 
and requires other energetic mechanisms driving the molecular gas. 

The velocity gradient along the loops and large velocity dispersions 
at their foot-points are typical features of solar magnetic loops 
emerging from  below the photosphere. 
The emergence of solar magnetic loops from the photosphere 
to the corona is driven by the 
buoyancy created by sliding the gas along the loop. 
Even when the unperturbed atmosphere is convectively stable, 
the magnetic loop can rise 
when the buoyancy at the loop top exceeds the restoring magnetic 
tension. 
This instability is called the Parker instability. 
Parker instability was originally proposed as a 
mechanism of the formation of the interstellar clouds  \citep{par1966}. 
\citet{mat1988} carried out two-dimensional magneto-hydrodynamic (MHD) 
simulations of the Parker instability in astrophysical disks 
and showed that in the nonlinear stage of the Parker instability, 
shock waves are formed near the foot-points of the magnetic loops 
because gas slides down supersonically(e.g., \cite{shi1989}). 
Such shocks can be an origin of large velocity dispersions 
observed near the foot-points of Galactic center molecular loops 
\citep{fuk2006}. 

Two-dimensional MHD simulations of the Parker instability in the 
Cartesian coordinate system of the gas disk can reproduce loop-like 
structures and velocity gradients similar to the observed loops 
\citep{fuk2006}. 
In their simulations, however, the Galactic rotation was not taken 
into account. 
In rotating stratified magnetized layer such as galactic gas disk, 
Coriolis force twists the magnetic loops. 
Since the twisting of magnetic fields enhances magnetic tension, 
the growth rate of the Parker instability is suppressed. 
\citet{cho1997} carried out local three-dimensional MHD 
simulations of the Parker instability by using a frame 
corotating with the disk 
to consider the effect of the Coriolis force.  
They confirmed that the Parker instability grows in rotating disks, 
although growth rate becomes smaller than that without rotation. 

In the gas disk rotating differentially, another instability, 
magneto-rotational instability (MRI) grows \citep{bal1991}. 
Three-dimensional local 3D MHD simulations adopting the shearing 
sheet approximation confirmed that MRI drives magnetic turbulence 
inside the disk (e.g., \cite{haw1995, mat1995, bra1995, 
sto1996, san1999}). 
When  we take into account the vertical gravity, 
buoyantly rising magnetic loops can be formed by the 
Parker instability. 
In the Galactic center region, 
since the loop size must be comparable to the 
distance from the Galactic center, non-local effects such as 
curvature becomes important. 
Global MHD simulations of the buoyant rise of magnetic loops from 
differentially rotating gas disks were first carried out by \citet{mac2000}. 
They showed that magnetic loops emerge from the turbulent disk. 
They assumed the gravitational force created by the 
central point mass. 
In Galactic gas disks, however, gravity is determined by 
stars and the dark matter. 

\citet{nis2006} presented the results of global three-dimensional MHD 
simulations of Galactic gaseous disks to study how the 
Galactic magnetic fields are amplified and maintained. 
They assumed a steady axisymmetric gravitational potential given 
by \citet{miy1975}. 
Their simulations showed that magnetic fields are amplified 
up to $\mu$ G and are maintained for 10 billion years. 
They also found that mean azimuthal magnetic fields reverse their 
direction in every 1Gyr. 
This field reversal is driven by the buoyant escape of magnetic 
flux from the disk to the Galactic halo. 
Their simulation region ($0.8 {\rm kpc} < \varpi < 30 {\rm kpc}$),  
however, did not include the galactic center. 
Here, $\varpi$ is the distance from the Galactic center. 

\citet{bae2008} proposed that collision of the high velocity cloud 
with the Galactic disk explains the formation of the 
molecular loop structure seen at the Galactic center, 
when the high velocity cloud has an oblique collision 
with the disk plane. 
They showed that the size was similar to that of the 
molecular loop, but the velocity gradient was smaller 
than that of the observation. 

In this paper, we report the results of global 3D MHD simulations of 
Galactic central gas disks 
inside 1kpc from the Galactic center. 
In section \ref{sec2}, we present basic equations and 
describe the numerical model. 
Numerical results will be presented in section \ref{sec3}. 
Section \ref{sec4} is devoted for discussion. 
Summary will be given in section \ref{sec5}.

\newpage

\section{Numerical Methods} \label{sec2}
\subsection{Basic equations}

We solved the following MHD equations in the cylindrical 
coordinate system $(\varpi, \varphi, z)$: 

%
%
\begin{equation}
       \frac{\partial \rho}{\partial t} 
    +  \nabla \cdot ( \rho \mbox{\boldmath $v$} )
  = 
       0 ~,
\label{eqn:b1}
\end{equation}
%
%
\begin{equation}
      \rho \left[
      \frac{\partial \mbox{\boldmath $v$}}{\partial t}
    + \mbox{\boldmath $v$} \cdot \nabla \mbox{\boldmath $v$}
      \right]
  = 
    - \nabla P 
    - \rho \nabla \phi
    + \frac{\mbox{\boldmath $j$} \times \mbox{\boldmath $B$}}{c} ~,
\label{eqn:b2}
\end{equation}
%
%
\begin{equation}
     \frac{\partial \mbox{\boldmath $B$}}{\partial t}
  = 
     \nabla \times 
    ( \mbox{\boldmath $v$} \times \mbox{\boldmath $B$} ) ~,
\label{eqn:b3}
\end{equation}
%
%
\begin{equation}
     \rho T \frac{d S}{dt}
  = 0 ~,  
\label{eqn:b4}
\end{equation}
where $\rho$, $P$, $\phi$, $\mbox{\boldmath $v$}$, $\mbox{\boldmath $B$}$, 
$\mbox{\boldmath $j$} = c \nabla \times \mbox{\boldmath $B$}/4 \pi$, 
$T$, and $S$ are the density, pressure, 
gravitational potential, velocity, magnetic field, current density, 
temperature and specific entropy, respectively. 
The specific entropy is expressed as 
$S = C_V {\rm ln} {(P/ \rho^{\gamma})}$,  
where $C_V$ is the specific heat capacity and 
$\gamma$ is the specific heat ratio. 

We adopt the Galactic gravitational potential proposed by 
\citet{miy1975}, 
\begin{equation}
     \phi(\varpi, z) 
 = -
  \Sigma_{i=1,2}  \frac{G M_{\rm i}}{[\varpi^2 + 
       \{a_{\rm i} +(z^2+b_{\rm i}^2)^{1/2} \}^2 ]^{1/2}} ~. 
\label{eqn:gr}
\end{equation}
%
%
Here, $G$ is the gravitational constant. 
The first term $(i=1)$ on the right hand side of equation (\ref{eqn:gr})
corresponds to the potential by bulge stars and 
the second term $(i=2)$corresponds to the disk component. 
The parameters adopted in this paper are shown 
in table \ref{tab2}. 

\begin{table}
 \caption{Parameters adopted for the gravitational potential}
 \label{tab2}
 \begin{center}
 \begin{tabular}{lccc} \hline \hline
            & $a_{\rm i}$ (kpc) & $b_{\rm i}$ (kpc) & 
      $M_{\rm i} ( 10^{10} M_{\odot}$) \\ \hline
  Bulge 1 & 0.0 & 0.495 & 2.05 \\
  Disk  2 & 7.258&0.520 & 25.47 \\ \hline 
 \end{tabular}
 \end{center}
\end{table}
    
\subsection{Numerical Methods and Boundary Conditions}

We solved the MHD equations 
by using a modified Lax--Wendroff scheme \citep{rub1967} with 
artificial viscosity \citep{rit1967}.  

For normalization, we use the units listed in table \ref{tab1}.
The units of length $r_0$ and velocity $v_0$ are 1 kpc and 
$v_0 = \sqrt[]{ G M_0 / r_0}$, respectively. 
Here, $M_0 =10^{10} M_{\odot}$.  
Thus the unit time is equal to $t_0=r_0 / v_0 = 4.7 \times 10^6 ~ {\rm year}$. 
The unit temperature is given by 
$T_0 = m_{\rm p} v_0^2 k_{\rm B}^{-1}= 5.2 \times 10^{6} ~ {\rm K}$, 
where $m_{\rm p}$ is the proton mass and 
$k_{\rm B}$ is the Boltzmann constant.
\begin{table}
 \caption{Units adopted in this paper.}
 \label{tab1}
\begin{center}
\begin{tabular}{lll} \hline \hline
Physical Quantity & Symbol & Numerical Unit \\ 
\hline 
Length & $r_0 $ &   1kpc \\
Velocity & $v_0$ & 207 km ${\rm sec}^{-1}$ \\
Time & $t_0$ & $ 4.7 \times 10^6$ year \\
Temperature & $T_0$ & $5.2 \times 10^{6}$ K \\ \hline
\end{tabular}
\end{center}
\end{table}

The number of grid points is 
$(N_{\varpi}, N_{\varphi}, N_z) = (200, 64, 384)$.  
The grid size is $\Delta \varpi = \Delta z = 0.01 $ for 
$0 < \varpi  < 1.2$ and $|z|<1.0$, and otherwise increases 
with $\varpi$ and $z$. 
The grid size in the azimuthal direction is $\Delta \varphi = 2 \pi / 63$. 
The outer boundaries at $\varpi = 11.58 $ and 
at $|z| = 16 $ are free boundaries where waves can be transmitted.  
We included the full circle $(0 \leq \varphi \leq 2 \pi)$ 
in the simulation region, and 
applied periodic boundary conditions in the azimuthal direction. 
An absorbing boundary condition is imposed  at 
$r = (\varpi^2+z^2)^{1/2} =  r_{\rm in} = 0.2 $ 
by introducing a damping parameter,  
\begin{equation}
       a_{\rm d} 
  =  
       0.1 \left(
       1.0 - \tanh{
       \frac{r - r_{\rm in} + 5 \Delta \varpi}{2 \Delta \varpi}}
       \right) ~.
\label{eqn:b6}
\end{equation}
The physical quantities 
$q = (\rho, \mbox{\boldmath $v$}, \mbox{\boldmath $B$}, P)$ 
inside $r = r_{\rm in}$ are re-evaluated by 
\begin{equation}
       q^{\rm new} 
 = 
       q - a_{\rm d} (q - q_0) ~,
\label{eqn:b7}
\end{equation}
which means that the deviation from initial values $q_0$ is 
artificially reduced with a rate $a_{\rm d}$. 
Thus waves propagating inside $r=r_{\rm in}$ are absorbed in the transition 
region ($r_{\rm in}-5 \Delta \varpi < r < r_{\rm in}$). 

A random perturbation is added in the azimuthal velocity 
at the initial state. 
The maximum amplitude of the perturbation is $1\%$ of the 
initial azimuthal velocity at the radius. 

\subsection{Initial Model}
At the initial state, we assume a gas torus 
around $\varpi \sim 1 {\rm kpc}$. 
The torus is threaded by a weak toroidal magnetic field. 
The initial torus is assumed to have a specific 
angular momentum, $L\propto \varpi^{a}$.  

According to \citet{oka1989},  
by using the polytropic relation $P=K \rho^{\gamma}$ 
at the initial state and by assuming 
$\beta = 8 \pi P/B_{\varpi}^2 = \beta_{\rm b} (\varpi/\varpi_{\rm b})^{1/2}$, 
%
%
where $\beta_{\rm b}$ is the initial plasma $\beta$ at the 
initial pressure maximum of the torus 
$(\varpi,z) = (\varpi_{\rm b},0)$, and $B_{\varphi}$ is the 
azimuthal magnetic field,  
we integrated the equation of motion into a potential form, 
\begin{equation}
      \Psi(\varpi,z) 
 = 
      \phi + \frac{L^2}{2 \varpi^2} + \frac{1}{\gamma-1}v_{\rm s}^2 
     + \frac{\gamma}{2(\gamma -1)}v_{\rm A}^2 = \Psi_{\rm b}  
 =  {\rm constant} ~ , 
\label{eqn:poten}
\end{equation}
where $v_{\rm s} = (\gamma P/\rho)^{1/2}$ is the sound speed, 
$v_{\rm A} = [B_{\varphi}^2/(4 \pi \rho)]^{1/2}$ is the Alfv{\'e}n speed, 
and $\Psi_{\rm b} = \Psi(\varpi_{\rm b},0)$.  
At $\varpi = \varpi_{\rm b}$, the rotation speed of the torus 
$L/\varpi_{\rm b}$ equals to the Keplerian velocity. 
By using equation (\ref{eqn:poten}), we obtain the density distribution as 
\begin{equation}
      \rho 
 = 
      \rho_{\rm b} \left\{
         \frac{\max{ [\Psi_{\rm b} - \phi -L^2/(2\varpi^2),0]}}
              {K[\gamma/(\gamma-1)]
                [1+\beta_{\rm b}^{-1}\varpi^{2(\gamma-1)}
       /\varpi_{\rm b}^{2(\gamma-1)}]} 
       \right\}^{1/(\gamma -1)}      ~ ,   
\end{equation}
where $\rho_{\rm b}$ is the density at 
$(\varpi,z) = (\varpi_{\rm b},0)$. 
Outside the torus, we assumed a hot, isothermal ($T=T_{\rm halo}$) 
spherical halo. 
The density distribution of the halo is given by 
$\rho_{\rm h} = \rho_{\rm halo} 
\exp [-(\phi-\phi_{\rm b})/(k_{\rm B} T_{\rm halo})]$, 
where $\phi_{\rm b}$ is the gravitational potential at 
$(\varpi,z)=(\varpi_{\rm b},0)$. 

In this paper, we assume 
that the initial gas ring (torus) consists of 
the warm component of the interstellar gas.  
If we assume the cold component ($T \sim 100 K$), 
the length of the magnetic loops will be much shorter than the 
length of observed loops. 
This situation is similar to the solar emerging magnetic loops. 
Their length is about 3000km when they emerge from the 
photosphere but the loops elongate as they rise. 
The elongated loop length eventually exceeds $10^4$ km 
in well-developed active regions. 
It means that their length is determined not by the 
photospheric temperature but rather by the coronal one. 
%
We do not include the effects of 
star formation, feedbacks to the gas dynamics by supernova explosions, 
nor winds from massive stars. 
The effects of the gas heating by supernova explosions are 
approximately taken into account by assuming warm interstellar gas 
whose scale height is about 150 pc at 1kpc from the galactic center. 
We ignore the self-gravity of the gas and the radiative cooling term. 
The sound speed assumed in this paper is 
$v_{\rm b} =0.14$ at $\varpi = \varpi_{\rm b}$ where the gravity 
equals to the centrifugal force. 
The initial temperature at $\varpi = \varpi_{\rm b}$ is 
$T_{\rm b} = 0.01 T_0$ ($ = 5.2 \times 10^4$ K). 
Model parameters adopted in this paper are $\varpi_{\rm b} = r_{0}$, 
$\beta_{\rm b} = 10$, $\gamma = 5/3$, $L=(\varpi_{\rm b}/2)^{1/2}
\varpi_{\rm b} /(\varpi_{\rm b}-1) \varpi^{a}$, $a=0.46$,  and 
$\rho_{\rm halo} = 10^{-5} \rho_{\rm b}$.


\section{Numerical Results} \label{sec3}
\subsection{Time Evolution of the Gaseous Disks}

The simulations show a general trend that the MHD instabilities 
take place over the entire disk and the gas is strongly driven 
by these instabilities. 
The gas temperature increases because of the dissipation of the 
magnetic energy in the disk. 

\begin{figure}
 \begin{center}
    \FigureFile(80mm,80mm){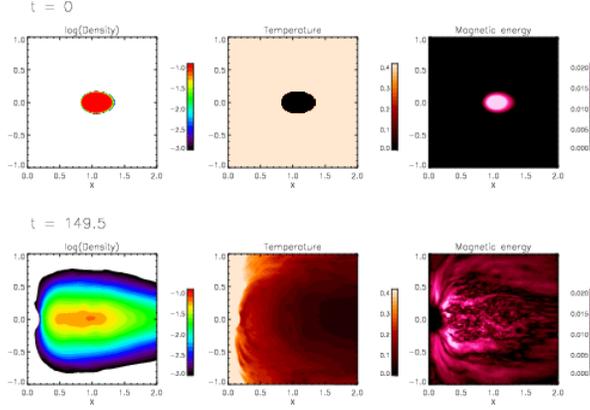}
 \end{center}
\caption{
Distribution of azimuthally averaged quantities for the initial state (top) 
and quasi-steady state at $t=149.5 = 0.7 {\rm Gyr}$ . 
Left panels show the logarithmistic density distribution. 
Middle panels show the azimuthally averaged temperature. 
Right panels show the magnetic energy.
\label{fig:ro2d}}
\end{figure}

Top and bottom panels of figure \ref{fig:ro2d} show the initial 
condition and the snapshot at $t=149.5$. 
Left panels show the density distributions averaged in the 
azimuthal direction.
Middle panels show the temperature distribution and 
right panels show the magnetic energy distribution.
Owing to the angular momentum redistribution, the initial 
torus is deformed into a thick disk. 
The disk temperature increases as the disk is heated by 
dissipation of the magnetic energy. 
%
%
When matter accretes in differentially rotating disks by 
transporting angular momentum outward, 
the released gravitational energy heats the disk gas. 
In magnetically turbulent disks, in which the angular momentum is 
mainly transported 
by Maxwell stress, the disk is heated by 
the magnetic energy dissipation (e.g., \cite{hir2006, mac2006}). 
The gas heating which takes place in our simulations also 
comes from this dissipation of the magnetic energy. 
Bottom right panel of figure \ref{fig:ro2d} shows that magnetic 
fields emerge from the disk and occupy the disk halo.

\begin{figure}
  \begin{center}
    \FigureFile(80mm,80mm){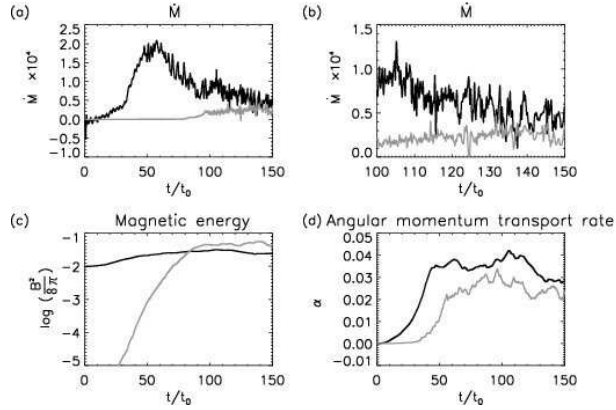}
  \end{center}
  \caption{(a) Time evolution of the mass accretion rate 
measured at $\varpi=0.8$ (black) and at $\varpi = 0.3$ (gray). 
Mass accretion rate is integrated in the azimuthal direction 
and in $|z| < 1$. 
(b) The same as (a) but for $100 < t/t_0 < 150$. 
(c) Time evolution of the magnetic energy 
averaged in the vertical direction ($|z|<0.5$), in the 
radial direction $0.8 < \varpi < 1.2$, and in the azimuthal 
direction (black curve). 
Gray curve shows the magnetic energy averaged 
in $0.3 < \varpi < 0.6$, $|z| < 0.1$ and in the azimuthal 
direction. 
(d) Time evolution of the averaged angular momentum transport rate 
$ \alpha \equiv - 
\langle B_{\varpi} B_{\varphi}/4 \pi \rangle/ \langle P \rangle$. 
Black curve shows that averaged in $0.8 < \varpi < 1.2$, and gray curve 
shows that averaged in $0.3 < \varpi < 0.6$. 
\label{fig:taccr}}
\end{figure}

Figure \ref{fig:taccr}a and \ref{fig:taccr}b show the time evolution 
of the mass accretion rate. 
Black and gray curves show $\dot{M}$ measured at 
$\varpi = 0.8$ and $\varpi = 0.3$, respectively. 
Mass accretion takes place quasi-steadily after $t=100t_0$. 
%
The mass accretion rate shows rapid time variations. 
Figure \ref{fig:taccr}c shows the time evolution of the 
azimuthally averaged magnetic energy. 
The black curve and the gray curve depicts the magnetic energy 
averaged over $|z|<5$ in the outer region ($0.8 < \varpi < 1.2$) 
and in the inner region ($0.3 < \varpi < 0.6$), respectively. 
At the initial state, the plasma $\beta$ in the outer region is  
$\beta \sim 10$. 
After a few dozens of rotations, the plasma $\beta$ in the outer region 
stays around $\beta \sim 10$. 
The right panels of figure \ref{fig:ro2d} shows that magnetic fields 
amplified in the disk are escaping from the initial disk, although 
the magnetic field is continuously amplified inside the disk 
by the MRI. 
%

At the initial state of our simulation, 
magnetic fields do not exist in the inner regions of the 
disk because we confined the azimuthal magnetic fields 
only inside the initial gas torus. 
As MRI grows in the outer disk, the disk gas infalling 
toward the central region carries in the magnetic fields. 
The magnetic fields are amplified by MRI within the time 
scale of several rotation period. 
The amplification of the magnetic fields saturates when 
$\beta \sim 10$, similarly to the results for accretion disks
(e.g.; \cite{mac2003}). 
Figure \ref{fig:ro2d} shows that the gas pressure in the 
galactic central region is higher than that for the outer region. 
Because of this higher gas pressure and faster rotation, 
magnetic fields in the inner region grow faster and stronger than 
those in the outer region.
Figure \ref{fig:taccr}d shows the time evolution of the 
angular momentum transport rate defined by $\alpha \equiv -  
\langle B_{\varpi}B_{\varphi}/4 \pi \rangle / \langle P \rangle$. 
Black and gray curves show $\alpha$  averaged in the same 
region as figure \ref{fig:taccr}c. 
The time averaged value of $\alpha$ is about $\alpha \sim 0.03$ 
in the outer region and $\alpha \sim 0.02$ in the inner region. 

\subsection{Structure of Magnetic Loops in the Coronal Region}


\begin{figure}
  \begin{center}
    \FigureFile(80mm,80mm){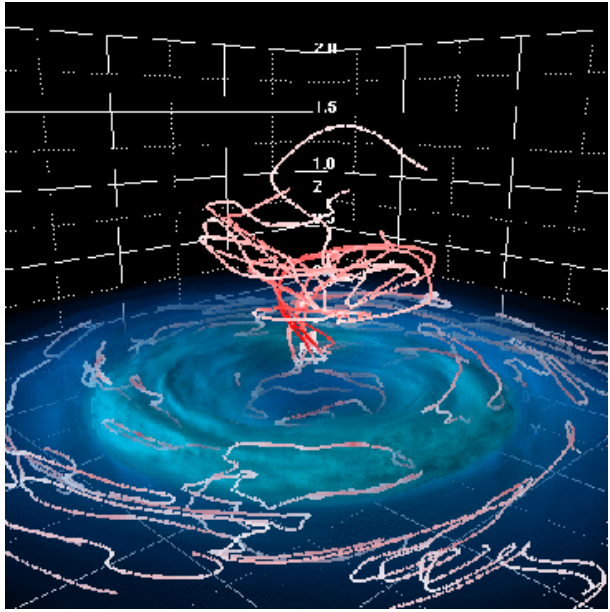}
  \end{center}
  \caption{Snapshot of the density distribution (blue color) 
and magnetic field lines at $t= 149.5$ ($\sim 0.7$ Gyr). 
Blue region shows the volume rendered density and 
curves show magnetic field lines.
\label{fig:vren}}
\end{figure}

Figure \ref{fig:vren} shows the density distribution and 3D structure 
of magnetic field lines. 
Blue surfaces show the volume rendered image of the density distribution. 
Bright color indicates the higher density region. 
Curves in figure \ref{fig:vren} show the magnetic field lines. 
This clearly shows that the magnetic fields become turbulent inside 
the gaseous disk. 
Since it is not easy to trace and identify individual magnetic field lines, 
we need a tool to extract field lines which belong to the 
same magnetic loop from the numerical results. 
We identified individual magnetic loops by the following algorithm. 

\begin{enumerate}
\item Choose a horizontal plane $S (z=z_0)$  which roughly 
coincides with the disk surface. 
\item Magnetic field lines starting from the surface $S$ 
are traced by integrating magnetic fields from each grid point on $S$. 
The integration is carried out by 4th-order Runge-Kutta method. 
Integration step length and maximum number of integration steps 
are $\Delta l$ and $N_{\rm s}$, respectively. 
Integration completes when the integrated field line crosses the 
original plane $S$.  
The start point $P_{\rm s}$ and the end point $P_{\rm f}$ 
reside on the same plane $S$. 
\item Pick up the magnetic loops which satisfy the following 
conditions; 
a) at the starting point, direction of magnetic field is 
toward the corona ($B_{z} > 0$), 
b) it crosses the original plane $S$ within $N_{\rm s}$ steps, 
c) the distance between the foot-points  is longer than $l_{\rm min}$, 
and d) the length of the magnetic field line $l$ is longer than $1$.  
\item Two magnetic field lines are identified as belonging to the 
same flux tube 
when the magnetic field lines starting from the 
point in the neiborhood of $P_{\rm s}$ also satisfies all 
the above conditions 
and the end point is within $2 \Delta x$ from  $P_{\rm f}$.  
Here $\Delta x$ is the grid size. 
\end{enumerate}

\begin{figure}
  \begin{center}
    \FigureFile(80mm,80mm){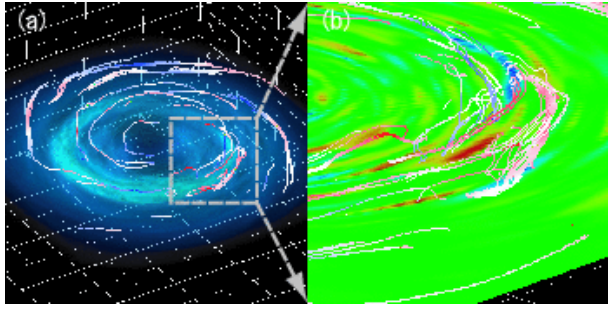}
  \end{center}
  \caption{(a) Volume rendered image of density (blue) and magnetic loops 
(colored curves). 
Colors on the field lines denote the vertical velocity 
($-0.05 < v_z < 0.05$ blue-white-red). 
(b) Zoom up image of magnetic field lines inside the gray box 
in (a). 
Colored slice shows the strength of the vertical magnetic field. 
Blue is negative and red is positive. 
\label{fig:mfl3d}}
\end{figure}

Curves in figure \ref{fig:mfl3d} show the examples of magnetic loops 
identified according to this algorithm. 
In this analysis, we chose $\Delta l = \Delta x = 0.02$, 
$z_0=0.1$, and $N_{\rm s}=200$. 
Figure \ref{fig:mfl3d} shows the density distribution and the 
magnetic loops in the region $-2 < x, y <2$, and $-1<z<1$. 
Blue surfaces in figure \ref{fig:mfl3d}a depict the volume rendered 
image of the density. 
Magnetic field lines are colored according to the vertical velocity 
($-0.05 < v_z < 0.05$, from blue to red). 
We can identify several loop like structures in the coronal region. 
Figure \ref{fig:mfl3d}b enlarges the region in 
dashed-gray box in figure \ref{fig:mfl3d}a. 
Colors in the green plane denote the direction of vertical magnetic field. 
Blue is negative and red is positive. 
It is clear that approximate height of the loop top is about $0.2$ kpc. 
Small loops continuously rise below the large magnetic loops.


\begin{figure}
  \begin{center}
    \FigureFile(80mm,80mm){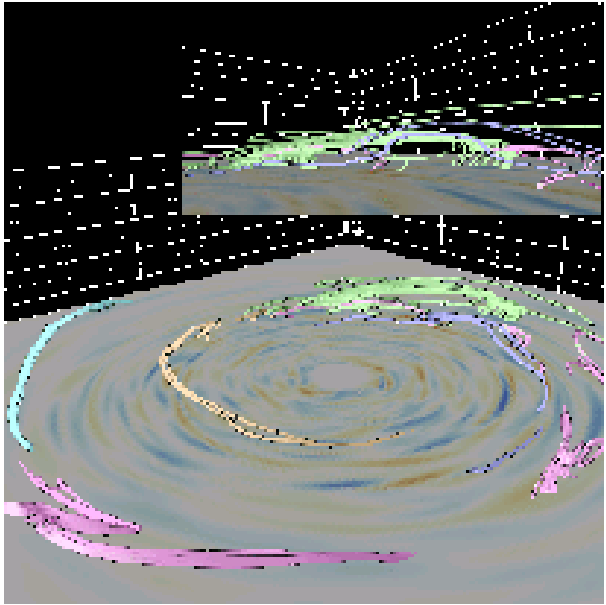}
  \end{center}
  \caption{Magnetic field lines in the coronal region. 
Magnetic field lines depicted by the same color show 
the magnetic field lines identified as 
belonging to the same flux tube. 
Slice locates at $z=0.1$. 
Color on the slice shows the 
magnetic energy.  
\label{fig:ftbz}}
\end{figure}

Figure \ref{fig:ftbz} shows the magnetic field lines 
selected by the same method as those in figure \ref{fig:mfl3d}. 
Color denotes the field lines identified as 
belonging to the same magnetic flux tube. 
Horizontal slice shows the magnetic energy distribution at $z=0.1$. 
The inset 
shows the same flux tubes projected from a different angle. 
When two or more magnetic field lines are included in the same flux tube  
(see green curves), we identify them as belonging to the same loop. 
The loop tops shift toward the outer radius 
due to the centrifugal force. 


\begin{figure}
  \begin{center}
    \FigureFile(80mm,80mm){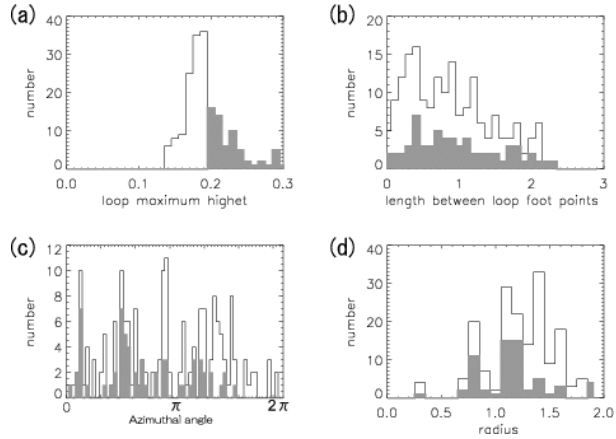}
  \end{center}
  \caption{Statistics of magnetic loops. 
(a) The histogram of the loop maximum height. 
(b) The histogram of the separation of loop foot points. 
(c) The azimuthal distribution of the loop top position.  
(d) The distribution of the radius of the loop tops. 
Gray shaded ones show the statistics of loops whose maximum height 
exceeds  $z=0.2$.  
\label{fig:hist}}
\end{figure}

Figure \ref{fig:hist} shows the statistics of the loops. 
Although a magnetic flux tube consist of a finite number of magnetic 
field lines, we chose a representative magnetic field line 
and let it represent the magnetic flux tube. 
For example,  maximum height of a loop is defined by 
the loop top of the representative magnetic field line. 
We identified 182 magnetic loops in the upper half of the 
gaseous disk at $t=149.5$.   
Figure \ref{fig:hist}a shows the number distribution of the loop top heights. 
Horizontal axis shows the loop height in unit of kpc. 
Gray shaded box shows the 
magnetic loops whose loop top exceed $z=0.2$. 
Integration of the magnetic field lines start from $z=0.1$ plane. 
We considered only the upper half above the equatorial plane ($z > 0$). 
Figure \ref{fig:hist}b shows the histogram of the loop lengths  
defined by the distance between the loop foot-points.  
Horizontal axis shows the separation of the foot-points in unit of kpc. 
Since two distributions in figure \ref{fig:hist}b are similar each other, 
the loop height and the loop foot-points distance 
do not have significant correlations. 

Figure \ref{fig:hist}c shows the azimuthal distribution of the 
number of magnetic loops. 
The magnetic loops whose loop top heights exceed $z=0.2$ 
concentrated in $\pi / 2 < \varphi < 3\pi / 2$. 
However, when we include the loops whose loop 
top is lower (white boxes), the tendency weakens. 
The radial distribution of the loop top is shown in figure \ref{fig:hist}d. 
Peaks appear around $\varpi \sim 0.8, 1.1$, and $1.4$. 
 

\begin{figure}
  \begin{center}
    \FigureFile(80mm,80mm){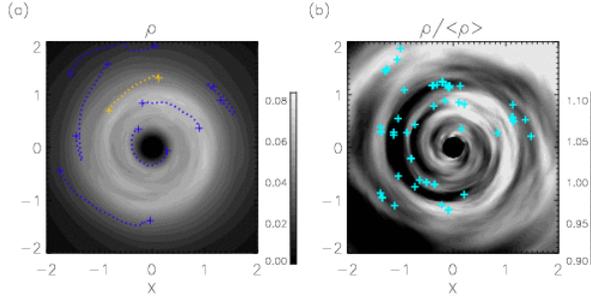}
  \end{center}
  \caption{(a) Equatorial density averaged in 
$|z|<0.06$. 
Gray scale indicates the density.
Dotted curves show the magnetic field lines projected onto the 
equatorial plane.
Symbols denote the position of foot-points of magnetic loops. 
(b) Equatorial density distribution normalized by the 
azimuthally averaged density $\rho / \langle \rho \rangle$. 
Crosses denote the position of the loop tops. 
\label{fig:roxy}}
\end{figure}

Next, let us show the distribution of loop tops projected onto 
the equatorial plane. 
Figure \ref{fig:roxy}a shows the
density distribution around the equatorial plane 
(averaged in $|z|<0.06$). 
Yellow and blue dotted curves show the same magnetic loops 
displayed in figure \ref{fig:ftbz} 
projected onto the equatorial plane. 
Yellow curves correspond to the green loops in figure \ref{fig:ftbz}. 
This figure shows that the density distribution has a $m=1$ 
one-armed distribution near the equatorial plane. 
Figure \ref{fig:roxy}b shows the 
normalized density$\equiv \rho/ \langle \rho \rangle$. 
Here, we defined the averaged density as follows. 
\begin{equation}
\langle \rho \rangle = \int_{-0.06}^{0.06} \int_0^{2 \pi} \rho d\varphi dz 
 / \int_{-0.06}^{0.06} \int_0^{2 \pi} d\varphi dz ~ .
\end{equation} 
In figure \ref{fig:roxy}b, 
crosses denote the position of the loop tops. 
It is shown the magnetic loops are deficient in $3\pi/2 < \varphi <2\pi$. 
This region corresponds to the higher density region. 
Loop tops are located near the lower density region, 
because the magnetic loops 
often buoyantly rise from the region where 
magnetic pressure becomes comparable to the gas pressure. 


\begin{figure}
  \begin{center}
    \FigureFile(80mm,80mm){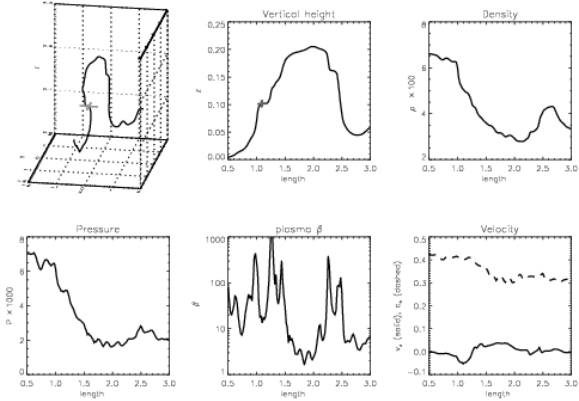}
  \end{center}
  \caption{Distribution of quantities along the loop depicted by 
yellow curve in figure \ref{fig:roxy}. 
Right bottom panel shows the vertical velocity 
(solid) and the sound speed (dashed). 
\label{fig:pval}}
\end{figure}

Figure \ref{fig:pval} plots the distribution of physical quantities 
along the magnetic field lines depicted by the yellow 
curves in figure \ref{fig:roxy}a. 
Horizontal axis shows the length along the magnetic field lines. 
Top rows are the 3D view of the magnetic loops, the vertical height, 
and density from left to right. 
Bottom rows are the distribution of the gas pressure, 
the plasma $\beta$,  vertical velocity and sound speed, respectively. 
These figures evidently show that magnetic fields are buoyantly rising.
That is, the density 
decreases from the foot points to the loop top. 
Plasma $\beta$ inside the loop is about $1$. 
Inside the magnetic loop, the sound speed is almost constant 
(see figure \ref{fig:pval}).  

\section{Discussion} \label{sec4}
\subsection{Loop Formation and the Origin of Their Concentration} 
As shown in figures \ref{fig:mfl3d} and \ref{fig:ftbz}, 
magnetic loops are formed in the halo region ($z>0.1$). 
We identified about 400 loops summing up the upper and lower 
coronal region. 
These magnetic loops are concentrated both 
in certain azimuthal range and radial range 
(figure \ref{fig:hist}c, and \ref{fig:hist}d). 
Magnetic fields inside the gaseous disk have turbulent component 
whose amplitude is comparable to the mean azimuthal magnetic fields. 
Therefore, the magnetic energy shows random patchy 
structure (see figure \ref{fig:ftbz}). 
On the other hand, one-armed non-axisymmetric pattern 
develops for density. 
Magnetic pressure in the low density region becomes 
stronger than that in the higher density region, 
because the total pressure is balanced in the disk.  
Since the magnetic loops with $\beta \sim 1$ are 
the most unstable to the Parker instability, 
buoyantly rising magnetic loops are formed in such lower density region. 
Thus, non-axisymmetric density distribution results in the 
azimuthal concentration of magnetic loops. 


\begin{figure}
  \begin{center}
    \FigureFile(80mm,80mm){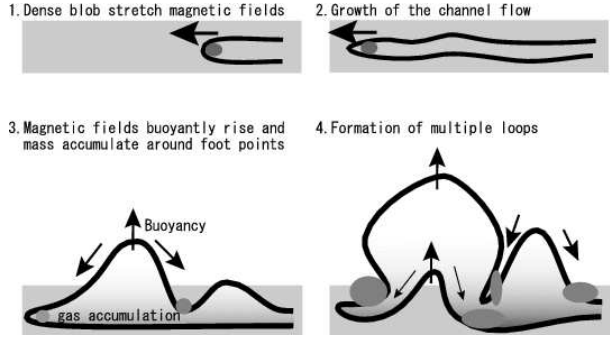}
  \end{center}
  \caption{Schematic picture showing the formation of 
magnetic loops. 
\label{fig:sch}}
\end{figure}

Figure \ref{fig:sch} shows schematically how magnetic loops are formed. 
As MRI grows, channel flows (e.g., \cite{haw1992, san2001, mac2003}) 
create large scale horizontal magnetic fields (figure \ref{fig:sch}b). 
When the magnetic energy accumulated in the channel flow becomes 
comparable to the thermal energy, 
magnetic loops emerge from the disk by buoyancy (figure \ref{fig:sch}c, d). 
This process creates chains of magnetic loops.

\subsection{Comparison with Observations}\label{sub:comp}
From their millimeter-wave CO observations,  \citet{fuk2006} 
found two molecular features that look like a loop with a 
length of several hundred parsecs and width of $\sim 30$ pc 
within $\sim 1$kpc from the center of our Galaxy. 
The kinetic energy of these CO loops is about $\sim 10^{51}$ erg. 
As shown in figure \ref{fig:ftbz} and  \ref{fig:hist},
our numerical simulation produced magnetic loops whose 
height $\sim 200$pc, width $\sim 50-300$pc, and 
length $\sim 0.1-2$kpc. 
To compare our simulation with observations, 
we assumed that the initial gas density is 
$\rho_{\rm b}= 1.6 \times 10^{-24} {\rm g/cm}^3$.  
In this case the total energy in a loop is about $10^{51}$ {\rm erg}. 


\begin{figure}
  \begin{center}
    \FigureFile(80mm,80mm){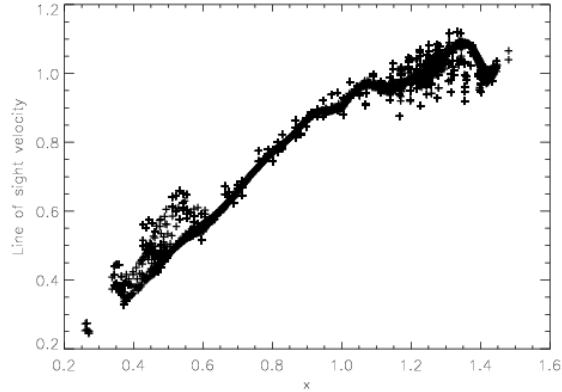}
  \end{center}
  \caption{Position-velocity diagram of the mass elements in 
the loops depicted in green in figure \ref{fig:ftbz}. 
This viewgraph shows the P-V diagram observed from the 
direction  $\theta = 3\pi /8 $. 
\label{fig:pvd} }
\end{figure}

In CO loops observed in the Galactic center,  
the line of sight velocity decreases linearly with the length along 
the loop (see fig.1 in \cite{fuk2006}). 
Figure \ref{fig:pvd} shows a  position-velocity diagram 
obtained by our simulation. 
We plotted magnetic loops colored in green in figure \ref{fig:ftbz}. 
The horizontal axis shows the $x$-direction in the Cartesian 
coordinate system and the vertical axis shows 
$y$-component of the velocity. 
We assumed that we observed the loops from infinity. 
Figure \ref{fig:pvd} shows the linear decrease of the line of 
sight velocity along the loop.  
Numerical results indicate that the rotation of the Galactic 
center gas disk creates this velocity gradient. 
From figure \ref{fig:pvd}, 
we can clearly see that foot points show the larger velocity 
dispersion than the loop top. 
It can be explained by the 
large velocity gradient around the foot-points of the 
magnetic loops where supersonic flow along the magnetic loops 
collides with the dense disk gas. 

Figure \ref{fig:vren} shows that a large-scale poloidal magnetic 
field is created in the innermost region of the Galactic gas disk. 
\citet{kat2004} showed that magnetic towers are formed by 
twisting of magnetic loops emerging from a magnetically turbulent 
accretion disk. 
Such magnetic towers can be formed even when the initial magnetic 
field is purely azimuthal \citep{mac2008}. 
The large scale poloidal magnetic fields penetrate the gas disk. 

\subsection{Mass Supply to the Central 100pc}

In our simulations, the unit of mass accretion rate is given by 
\begin{equation}
\dot{M}_0  = 33 \left( \frac{\varpi}{1{\rm kpc}} \right)^2
                \left( \frac{v}{207 {\rm km/s}}  \right)
                \left( \frac{\rho}
                      {1.66\times 10^{-24} {\rm g/cm^3}} \right) 
               ~ M_{\odot} /{\rm year} ~ .
\end{equation}
According to figure \ref{fig:taccr}b, 
mass accretion rate at $200 {\rm pc}$ in a quasi-steady state is 
$\dot{M} \sim 10^{-5} \dot{M}_0  \sim 3.3 \times 
10^{-4} M_{\odot} / {\rm yr}$. 
This mass accretion rate is determined not by the 
mass of the central black hole but by the mass of the interstellar gas. 
Note that the initial density of the 1kpc gas ring 
we assumed as an initial condition 
is arbitrary in our simulation. 
Thus, when dense gas infalls in this region, 
the accretion rate can be much higher. 

The mass accretion rate near the event horizon of the 
Galactic central black hole ${\rm SgrA^*}$  is estimated to be less than 
$ \leq 10^{-8} M_{\odot} {\rm yr}^{-1}$, assuming an 
equipartition magnetic field (e.g., \cite{bag2003}). 
Our simulation indicates the mass accretion rate at 
$ \varpi = 200$ pc becomes about 
$\simeq 10^{-4} M_{\odot} {\rm yr}^{-1}$. 
This may show that the mass accretion rate decreases from 
$10^{-4} M_{\odot} {\rm /yr}$ to $10^{-8}M_{\odot}/ {\rm yr}$ in 
$0.01 {\rm pc} < \varpi < 200 {\rm pc}$ by some 
mechanism such as star formation, supernova explosions, 
or by mass outflows.

\subsection{Effects of Cooling}
In this paper, we ignored the cooling of the gas. 
Therefore, the gas temperature of the gas disk exceeds $10^5 {\rm K}$. 
Since this temperature is higher than that of the warm component 
observed in the Galactic gaseous disk,  
the disk scale height 
may be overestimated. 
Since the most unstable wavelength of the Parker instability 
is  equal to 10 times of the disk scale height, 
the loop length in our simulation 
can be longer than that of the real galaxy. 
When we consider the radiative cooling, 
the interstellar gas will decompose into cold phase and the warm phase. 
Dense molecular clouds will be formed in the foot-points of 
magnetic loops where interstellar gas accumulates. 

In section \ref{sub:comp}, we compared the position-velocity diagram 
expected from our simulation with the observation. 
The velocity dispersion at the foot-points is 
smaller in our simulations than observations. 
Since the gas temperature and sound speed are overestimated
in our simulations, 
shock waves formed at the loop foot points are weaker 
than that expected for simulations including cooling. 
Thus, when we include the cooling, 
the position-velocity diagram may indicate a large jump 
similar to that observed in the Galactic center molecular loops.

\section{Summary} \label{sec5}

In this paper, we showed that when interstellar gas is supplied to 
the Galactic central region, magnetic loops are formed due to 
the buoyant rise of magnetic fields amplified by MRI inside the disk. 
When the mass of the supplied gas is about $10^7 M_{\odot}$, 
the magnetic fields can be amplified up to 10$\mu$G at 1kpc 
from the Galactic center. 
This enables the accretion of the ISM. 
Mass accretion rate at quasi-steady state is 
$\dot{M} \sim 10^{-5}-10^{-4} M_{\odot} {\rm /yr}$. 
Magnetic loops with length 1kpc 
buoyantly emerge from the disk 
and form magnetic loops whose maximum height exceeds 200pc. 
The disk gas slides down along the loop with 
maximum speed of 30km/s. 
We developed an algorithm which identify magnetic loops from 
simulation results. 
More than 400 loops are identified by applying this algorithm. 
Typical energy of a magnetic loops is $10^{51}$ erg. 
This magnetic energy is consistent with the kinetic energy of 
the CO loops observed by NANTEN telescope. 
We found that magnetic loops have preferential azimuthal angle. 
They tend to rise in regions with 
higher magnetic pressure (i.e., lower density). 

In this calculation, we considered only the warm gas components. 
However, the Galactic gaseous disk has cooler components. 
In subsequent papers, we would like to include the gas cooling effects.

\bigskip
We are grateful to Drs. T. Kudoh, K. Wada and 
K. Asano for useful discussion. 
Numerical computations were carried out on VPP5000 at 
Center for Computational Astrophysics, CfCA of NAOJ (P.I. MM). 
This work is financially supported in part by a Grant-in-Aid for Scientific
Research (KAKENHI) from JSPS(P.I. YF:20244014). 


\end{document}